\documentstyle[12pt] {article}
\begin {document}
\parindent=15pt
\begin{center}
\vskip 1.5 truecm
{\bf THE $x$-REGION OF SHADOWING CORRECTIONS
IN NUCLEON STRUCTURE FUNCTIONS}\\
\vspace{.5cm}
Yu.M.Shabelski \\
\vspace{.5cm}
Petersburg Nuclear Physics Institute, \\
Gatchina, St.Petersburg 188350 Russia \\
and \\
D.Treleani \\
\vspace{.5cm}
Univ. di Trieste; Dpt. di Fisica Teorica \\
Strada Costiera 11, Miramare-Grignano, I-34014 Trieste, Italy \\
\end{center}
\vspace{1cm}
\begin{abstract}
We discuss the experimental indications on the behaviour of $F_2(x,Q^2)$
at small
$x$ both in proton and nuclear targets. By comparing the
parametrizations of the data we conclude that shadowing correction effects in a
proton target can appear at a noticeable level for $x = (2 \div 4) \times
10^{-4}$ and $Q^2 \sim 10^1$ GeV$^2$, namely inside the HERA regime.

\end{abstract}
\vspace{3cm}

E-mail shabel@vxdesy.desy.de \\

E-mail daniel@trieste.infn.it \\

\newpage

\section{Introduction}

The experimental data of HERA on $F_2(x,Q^2)$ at small $x$, actually
$10^{-4} < x < 10^{-2}$, show at moderate $Q^2$ (say $Q^2 \sim 10^1$
GeV$^2$) a singular $x$-behaviour. Both ZEUS \cite{ZEUS} as well as H1
\cite{H1} Coll. data can be paramertrized as $x^{\delta}$ with
$\delta = -0.1 \div -0.25$. Evidently such a behaviour at $x \rightarrow 0$
is in contradiction with unitarity and it should be stopped by some
shadowing mechanism \cite{GLR,MQ,KMRS}. Although 
the estimations of the $x$-region where
this shadowing should appear are model dependent more or less
realistic calculations \cite{AGR,BGNPZ} predict significant (of the order
of ten percent) shadowing corrections at $x > 10^{-4}$. Moreover, in
accordance with \cite{AGL} the discussed singular behaviour is in
contradiction with unitarity at $x \sim (2\div 3)\times 10^{-4}$.

However no shadowing effect is actually seen in the experimental data on
$F_2(x,Q^2)$ obtained from proton targets at HERA. On the other hand the
singularity of $F_2(x,Q^2)$ at $x < 10^{-2}$ observed in nuclear
targets is significantly weaker, or, possibly, absent,
when compared with the singularity of $F_2(x,Q^2)$ observed on protons. It seems
therefore most likely that in the case of a proton target the parton 
density, in the
$x$ region investigated till now, is not high enough to have shadowing,
which is rather seen in
nuclear targets as a consequence of the enhanced parton density in a nucleus 
(say, by factor $\sim A^{1/3}$). In this connection we
present a new estimate of the $x$-region where shadowing corrections can be
significant in a proton target. Our estimate is not based on
specific model
assumptions, it is rather obtained by comparing the behaviour $F_2(x,Q^2)$  on
proton and nuclear targets.

\section{Discussion of experimental data from proton and
nuclear targets}

Let us consider the experimental situation more detail. The H1 Coll. presents
a global fit of their data of the singlet quark distribution,
$q_{SI} = u +\overline{u} + d +\overline{d} + s + \overline{s}$, which
determines practically the $F_2(x,Q^2)$  behaviour at small $x$ in the
form \cite {H1}
\begin{equation}
xq_{SI}(x) = A x^B(1-x)^C(1+Dx)
\end{equation}
with $A = 1.15, B = -0.11, C = 3.10$ and $D = 3.12$, at $Q^2$ = 4 GeV$^2$
and $x > 2 \cdot 10^{-4}$.  One can see that, at $x \leq 10^{-2}$, one can
rewrite this expression as
\begin{equation}
xq_{SI}(x) = A x^B
\end{equation}
and one may neglect the non-singlet contribution within a few percent 
accuracy.

ZEUS Collaboration presented their data for singlet quark distribution in
a similar form \cite{ZEUS}
\begin{equation}
xq_{SI}(x) = A x^B(1-x)^C(1+D\sqrt{x}+Ex)
\end{equation}
with $A = 0.47, B = -0.26, C = 3.29, D = 2.45$ and $E = 6.38$, at $Q^2$ =
7 GeV$^2$ and $x > (2 \div 3) \times 10^{-4}$. Also in this case one can rewrite
the expression in the form in Eq.(2). One may point out
a possible inconsistency in the two determinations of the value of $B$,
whose difference seems to be too large to be justified by
the difference in the values of $Q^2$.

As already  mentioned, the $x$ dependences of $F_2(x,Q^2)$ in nuclear
targets is not so singular. Usually the ratio
\begin{equation}
R_A(x) = F_2^{(A)}(x)/F_2^{(D)}(x)
\end{equation}
is considered, where $F_2^{(A)}(x)$ and $F_2^{(D)}(x)$ are the structure
functions per nucleon in a nucleus with mass number $A$ and in the deuteron
respectively.
The distortion of parton distributions by nuclear medium leads to
a deviation of $R_A(x)$ from unity\footnote{It is assumed usually that
the nucleons inside a deuteron are practically free.}. At small $x$ the
ratio can be expressed, independently of $Q^2$, as\cite{Smi}
\begin{equation}
R_A(x) = C \cdot x^{\gamma}\;,
\end{equation}
where $\gamma$ increases with $A$ for light nuclei and possibly becomes a
constant for $A \simeq (40-60)$. By fitting the experimental data in
\cite{Smi} one obtains the values $\gamma = 0.0874\pm 0.0037$, $0.0920\pm 0.0125$ and
$0.119\pm 0.020$ for $Ca, Cu$ and $Pb$, respectively. The values are
in agreement with the older analysis of \cite{Kari}, where $\gamma \sim
0.1$ was obtained for gold at $Q^2$ = 4 GeV$^2$ with weak
$Q^2$-dependence of $\gamma$.  One can therefore see that the
values of $F_2^{(A)}(x,Q^2)$ are practically independent on $x$
in heavy nuclei, if one uses the
H1 data for nucleon target, actually Eq.(1), and the singularity is about two
times weaker if one uses ZEUS data, namely eq.(3). In both cases one concludes
that shadowing of parton distributions in heavy nuclei is a
significant effect.

\section{Possible connection between shadowing effects in a nucleon and in a
nucleus}

Our main assumption to establish
a possible connection between the parton distributions
in nuclei and in free nucleons is that the
parton distributions at $Q^2 >> 1$ GeV$^2$ are determined by hard
processes which, in principle, can be described by QCD (possibly with some
non-perturbative contribution). In such a case all nuclear medium effects
should also be determined by hard parton interactions and not by some
soft physics (say, interactions with pion fields, etc.). As
a consequence all shadowing
effects in nuclei should be connected with the possibility of interacting
with partons from different nucleons, which effectively increases the parton
density. The highest enhancement of the parton density in a nucleus
is therefore $\sim A^{1/3}$, when all parton at some given
impact parameter can be involved in the interaction.

By parametrizing the data on $R_A(x)$ 
in the range 0.5 GeV$^2 < Q^2 < 200$ GeV$^2$ 
as in Eq.(5), one finds that
for comparatively heavy nuclei (say, for $Cu$) the values of the parameters
in Eq.(5) are $C \approx 1.3$ and $\gamma \approx 0.1$. One obtains
therefore
$R_A(x) = 1$ at $x \approx 0.06$ and the ratio 
decreases to the value $R_A(x_1) =
0.9$ (10\% shadowing effect) at $x_1 \approx 0.025$.  Following the
modern sets of phenomenological parton distributions \cite{Plo} one can
see that the value of $F_2^{(N)}(x_1)$, for a proton target at $x_1
\approx 0.025$ and $Q^2 \sim 10^1$ GeV$^2$, is about 0.45. In the case
of copper, where shadowing is observed, the parton density can be enhanced
by a factor as big as $A^{1/3} = 4$. 
It seems therefore quite reasonable to expect
the same shadowing effect for a proton target in the $x$-region where
$F_2^{(N)}(x_2) = 4\times F_2^{(N)}(x_1) \sim 1.8$. From the
review of parton distributions \cite{Plo} one can see 
that the correspondent $x_2$
value is $x_2 \sim 3\times 10^{-4}$, which is inside the kinematical
domain accessible at 
HERA\footnote{If we do the same estimate from the
parametrization of heavier nuclei we obtain smaller values of
$x_2$ because the values of parameters $C$ and $\gamma$ are practically
the same \cite{Smi} but the value of $A^{1/3}$ is larger. Possibly, 
in this kinematical regime, only a fraction of all partons at a given impact
parameter are allowed to interact in heavier nuclei.}.

A similar estimate can be obtained from the
parametrization of $R_A(x)$ in Eqs.(36) and (37) of Ref. \cite{WG}:
\begin{eqnarray}
R_A(x) & = & 1 + 1.19 [x^3-1.5(x_0+X_L)x^2+3x_0x_Lx] \ln^{1/6}(A)
\nonumber \\
& - & \left [\alpha_A - \frac{1.08(A^{1/3}-1)}{\ln (A+1)} \sqrt{x}\right ]
\exp{-x^2/x_0^2}
\end{eqnarray}
with $x_0 = 0.1$, $x_L = 0.7$ and $\alpha_A = 0.1 (A^{1/3} - 1)$.
The main difference between this parametrization and Eq.(5) is the behaviour
at $x \rightarrow 0$. Eq.(5) presents the power behaviour, whereas Eq.(6) 
gives a constant value at $x \leq 10^{-4}$. If one
looks for the value of $x$ where $R_A(x_1) = 0.9$ one obtains $x_1\sim
0.04$. The value of $F_2^{(N)}(x_1)$ is about 0.4 which gives
$F_2^{(N)}(x_2) = 4\times F_2^{(N)}(x_1) \sim 1.6$, corresponding to the
value $x_2\simeq4 \times 10^{-4}$.

\section {Conclusion}

It seems quite probable that shadowing corrections can be seen 
at the smallest values of $x$ accessible at HERA as
a deviation of $F_2(x,Q^2)$ from the behaviour in Eq.(2).

Two are the main sources of uncertainty in our estimate. 
The first is the quality of DIS data 
on proton and nuclear targets, which are used
to obtain the parameters in Eq.(1) and in Eq.(3).
The second is the enhancement factor. Actually $A^{1/3}$
has to be understood only as an order of magnitude estimate.
The real enhancement factor is
determined by the number of target nucleons which can interact with the wee
partons emitted by the virtual photon. It seems nevertheless rather
plausible that, by increasing the precision of the experimental
measurements of $F_2(x,Q^2)$ at the smallest values of $x$ available
at HERA, the first evidence of shadowing in proton structure
functions might be actually observed.
\vskip.25in

{\bf Acknowledgements}

\vskip.25in
We are grateful to M.G.Ryskin and G.I.Smirnov for useful discussions.

This work is supported in part by INTAS grant 93-0079.

\newpage



\end{document}